%% file: main.tex
\documentclass[9pt,twocolumn,twoside]{opticajnl}
\journal{opticajournal} 

\setboolean{shortarticle}{true}

\newboolean{arxivpreprint}
\setboolean{arxivpreprint}{true}
\ifthenelse{\boolean{arxivpreprint}}{}{\linenumbers} 

\title{Foundry-Fabricated Silicon Nitride Grating Arrays with Uniform Power Distribution at Visible Wavelengths}

\author[1,*]{Amged~Alquliah}
\author[1]{Xiaoxin~Wang}
\author[1]{Jifeng~Liu}
\author[1,$\dagger$]{Anthony~Rizzo}

\affil[1]{Thayer School of Engineering, Dartmouth College, Hanover, NH 03755, USA}

\affil[*]{amged.alquliah@dartmouth.edu}
\affil[$\dagger$]{anthony.j.rizzo@dartmouth.edu}

\begin{abstract}
Multiplexed optical biosensors are widely deployed to read out many biological targets in parallel. However, their channel count and integration density are limited by the photonic network that delivers light to individual sensing sites. Existing on-chip networks have been largely developed for phased arrays and image projection, where performance is assessed through an aggregate radiation pattern rather than the local intensity at each output. In this Letter,
we demonstrate a novel bus-and-tap network architecture for massively parallel multiplexed optical delivery. The network uses cascaded directional couplers with individually designed interaction lengths, allowing for arbitrary, user-defined illumination patterns not possible with conventional binary tree architectures. We experimentally validated our proposed architecture with 46 visible-wavelength grating emitter channels designed for uniform illumination at 40 $\mu$m pitch in a commercial 300-mm silicon nitride foundry process, achieving an emitted-intensity coefficient of variation of 9.3\% at 530~nm, and 8.8\% under band-integrated illumination from 550~nm to 650~nm. These results provide a route to scalable visible-light delivery for sensing, imaging, neural interfaces, and parallel optical addressing in quantum systems.
\end{abstract}

\setboolean{displaycopyright}{false} 
\doi{} 

\begin{document}

\maketitle

\section{Introduction}

Wearable and point-of-care diagnostics increasingly depend on
multiplexed biosensors that read out many biomolecular targets
simultaneously. These devices distribute light across an array of sensing
sites, each read out by a separate detector
\cite{werley2020multiplexed, miller2026multiplexed}. Scalable multiplexing
therefore depends not only on the sensing chemistry and detectors, but also
on the photonic network used to illuminate the sensing sites. Such a network must satisfy several requirements. First, the network should ideally operate in the visible band, since the
520--650~nm range supports widely used fluorescence- and
Raman-based biosensing schemes \cite{hang2022visible, tyndall2024foundry}, as well as the high-responsivity region of back-illuminated CMOS image
sensors \cite{manickam2017fully}. Second, it must deliver uniform
power because channel-to-channel intensity variation produces unequal signal levels and complicates calibration across the array
\cite{montero2021best}. Its routing should also remain practical as the
number of sensing sites increases. Visible-light silicon nitride
(Si$_3$N$_4$) photonics satisfies the wavelength requirement by supporting
low-loss propagation across the visible band in commercial CMOS foundry
processes \cite{poon2024silicon}.

The commonly adopted power distribution topology is the binary splitter tree, whose power split ratio is ideally symmetric. This topology has previously been applied to large-scale Si$_3$N$_4$ emitter arrays at infrared and visible bands \cite{poulton2016large}. However, generating $N$ outputs requires $N-1$ splitters, with at least $\lceil\log_2 N\rceil$ stages along the distribution path. Hence, excess loss and fabrication-induced imbalance can accumulate through successive stages \cite{zhang2023impact}. Moreover, a binary tree expands as a two-dimensional branching network, whereas the emitters are arranged along a one-dimensional line. Therefore, routing the tree outputs to the linear detector arrays requires additional footprint and complexity, introducing bends, possible waveguide crossings, and path-length matching constraints. Channel counts that are not powers of two, such as $N=46$, also do not map naturally onto a complete balanced tree of identical symmetric splitters and may require unused outputs, unequal split ratios,
or unequal-depth branches.

Serial bus-and-tap networks provide a straightforward architecture for linear emitter arrays. A single bus waveguide distributes optical power to the outputs through a cascade of directional couplers (DC), with each coupler designed to extract an independently specified fraction of the remaining power. Because the taps follow the same linear arrangement as the emitters, each tap can be placed adjacent to the emitter it feeds, thus eliminating the need for a separate fan-out network. The coupling coefficient can also be selected to accommodate arbitrary channel counts and power-distribution profiles. Such architectures have been demonstrated at 1550~nm
\cite{poulton2017coherent}, in passive $32\times32$ visible-light
image-projection arrays operating at 635~nm \cite{raval2018integrated}, in liquid-crystal-tunable visible-light taps \cite{notaros2024integrated}, and in $10\times100$ photonic routing manifolds with individually calibrated beam taps
\cite{chiles2018design}. These systems were designed for coherent beam
forming, image projection, or photonic routing, and their reported
figures of merit emphasize collective near- or far-field patterns rather
than per-output intensity uniformity.

A detector-matched biosensor front end presents a different
and unique design challenge. Each emitter illuminates its own
sensing and detector sites according to a pitch determined by
the detector and microfluidic designs \cite{werley2020multiplexed}. Therefore, the intensity
delivered to each independent site is the relevant metric rather than
the collective illumination pattern. Unlike coherent
beam-forming networks, the outputs are used independently and do
not require controlled relative phase among the branches. To
the best of our knowledge, there is no previous report on
per-emitter intensity uniformity for a large passive serial
bus-and-tap network operating at visible wavelengths.

Here, we demonstrate a passive bus-and-tap network feeding a
$1\times 46$ linear grating-emitter array at a 40-$\mu$m pitch, fabricated
in a commercial 300-mm foundry Si$_3$N$_4$ platform. We employ individually-designed variable DC interaction lengths to compensate for input power depletion across the bus waveguide. Additionally, the circuit was characterized as-fabricated, without trimming or active tuning. This passive equalization avoids per-channel heaters and calibration, reducing the overall system complexity. Narrowband profiling yields an emitted-intensity coefficient of variation (CV) of 9.3\% at 530~nm over a 15-emitter field of view, and the measured spatial gradient differs from the monotonic depletion predicted for a matched constant-tap network. A band-integrated measurement from 550 to 650~nm yields a CV of 8.8\%. Our demonstrated architecture therefore provides a scalable route to uniform visible-light distribution across CMOS-matched biosensor arrays, and extends naturally to parallel light-delivery networks for neural interfaces \cite{mohanty2020reconfigurable} and atom-based quantum systems, including trapped ions \cite{
niffenegger2020integrated} and neutral atoms
\cite{bluvstein2022quantum}.

\begin{figure}[ht!]
\centering
\includegraphics[width=\columnwidth,keepaspectratio]{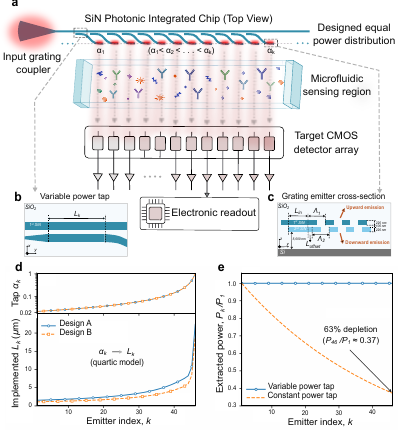}
\caption{\textbf{a}, Architecture: a focusing grating coupler feeds a bus with 46
cascaded DCs of increasing tap ratio $\alpha_k$, each driving a dual-layer grating
emitter; microfluidics and CMOS detectors indicate future target integration and are
not demonstrated here. \textbf{b}, Variable power tap with interaction length
$L_k$. \textbf{c}, Dual-layer emitter cross-section showing $L_\mathrm{in}$,
$\Lambda_1$, $\Lambda_2$, and $L_\mathrm{offset}$; design after
ref.~\cite{garcia2024visible}, with details in Supplement~1. \textbf{d}, Target
$\alpha_k$ and implemented $L_k$ for Designs~A and B. \textbf{e}, Ideal lossless
$P_k/P_1$ for variable and constant taps
($\alpha_c=\alpha_1$, 63\% depletion).}
\label{fig:concept}
\end{figure}

\section{Design and Fabrication}

The circuit uses the serial bus-and-tap architecture shown in
Fig.~\ref{fig:concept}a. A focusing grating coupler launches light into a main
Si$_3$N$_4$ bus waveguide. Along the bus, 46 cascaded DCs extract prescribed fractions of the remaining optical power into individual
branches. Each branch terminates in an out-of-plane grating emitter based on the
design approach reported in ref. \cite{garcia2024visible}, with the emitters placed at a 40-$\mu$m pitch spanning a length of 1.8~mm. Emitter geometry and simulations are described in detail in Supplement 1. A second grating coupler collects the residual bus power after the final tap.

For an ideal lossless bus, branch $k$ receives
\begin{equation}
P_k=\alpha_k P_{\mathrm{rem},k}=T,
\label{eq:tap}
\end{equation}
where $P_{\mathrm{rem},k}$ is the power arriving at the $k$\textsuperscript{th} tap,
$\alpha_k$ is its power-coupling coefficient, and $T$ is the target power per
branch. Because each preceding tap removes the same absolute power,
$P_{\mathrm{rem},k}=P_0-(k-1)T$, where $P_0$ is the power before the first tap.
The required coupling coefficient for the $k$\textsuperscript{th} tap is therefore
\begin{equation}
\alpha_k=\frac{T}{P_0-(k-1)T}\quad,
\label{eq:schedule}
\end{equation}
so the coupling coefficients increase along the array to compensate for bus
depletion, with $\alpha_1=T/P_0$. More generally, an arbitrary target power profile can also be realized by calculating the corresponding sequence of tap coefficients.

For the equal-power design considered here, complete extraction would require
$\alpha_N=1$. To keep the final tap within the simulated monotonic coupling
range, we instead set $\alpha_N=\alpha_{\max}<1$, which fixes the per-branch
target at
\begin{equation}
\frac{T}{P_0}
=
\frac{\alpha_{\max}}
{1+(N-1)\alpha_{\max}}.
\label{eq:target}
\end{equation}

We implemented two network variants on the same die, with Designs~A and B targeting nominal wavelengths of 550 and 650~nm, respectively. For each design, the DC response was simulated at its nominal wavelength using Ansys Lumerical
FDTD with fundamental TE-mode excitation and a 200-nm coupling gap. The simulated interaction length was fitted as a fourth-order polynomial of the coupling coefficient and used to map each target $\alpha_k$ to $L_k=L(\alpha_k)$ (Fig.~\ref{fig:concept}b). For $N=46$, the selected maximum
coupling coefficient gives $T/P_0\approx0.0217$ for both designs. The implemented interaction lengths span 1.55--22.41~$\mu$m for Design~A and 1.11--18.87~$\mu$m for Design~B; the complete tap details are provided in Table~S2 (Supplement 1).

In the ideal lossless model, the unequal-tap network gives $P_k/P_1=1$ across
all 46 branches (Fig.~\ref{fig:concept}c). A constant-tap network with
$\alpha_{\mathrm{c}}=\alpha_1$ instead gives
\begin{equation}
\frac{P_k}{P_1}=(1-\alpha_{\mathrm{c}})^{k-1}.
\label{eq:consttap}
\end{equation}
For $\alpha_{\mathrm{c}}=0.0217$, the final branch receives
$P_{46}/P_1=0.37$, corresponding to a 63\% decrease across the array. The
present network assumes a lossless bus. Inter-tap propagation and
DC excess losses could be incorporated in future designs through a recursive tap approach based on the bus transmission between
consecutive couplers.

The layouts were generated in GDSFactory using the AIM Photonics Nitride-Only process design kit (PDK). The bus, tap branches, and DCs use the upper 220-nm-thick Si$_3$N$_4$ layer, with nominal widths of 380~nm for Design~A and 500~nm for Design~B, and a 200-nm coupling gap. Each tap connects through an S-bend and taper to a 1.0-$\mu$m-wide dual-layer grating emitter. The upper and lower Si$_3$N$_4$ layers are each 220~nm thick and separated by 100~nm of SiO$_2$. 

\begin{figure}[ht!]
\centering
\includegraphics[width=\columnwidth,keepaspectratio]{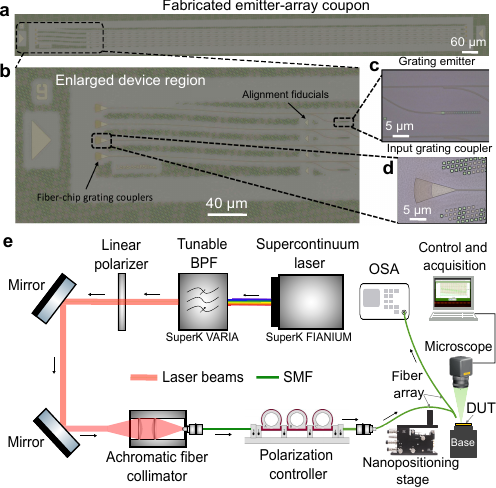}
\caption{Fabricated device and measurement setup. \textbf{a}, Optical micrograph of the emitter-array chip. \textbf{b}, Enlarged device region showing the input coupler, bus waveguide, emitter array, and alignment fiducials for future CMOS integration. \textbf{c}, Individual grating emitter micrograph. \textbf{d}, Focusing input grating coupler micrograph. \textbf{e}, Experimental setup for wavelength-filtered excitation and camera imaging. BPF: band pass filter; OSA: optical spectrum analyzer; DUT: device under test.}
\label{fig:device}
\end{figure}

For Design~A, the upper- and lower-layer grating periods are 0.72 and
0.75~$\mu$m, respectively, with a fill factor of 0.51 and a lower-layer
longitudinal offset of 0.36~$\mu$m. Design~B uses a 0.65-$\mu$m period in both
layers, a fill factor of 0.50, and an offset of 0.88~$\mu$m. The
grating-emitter input-taper lengths are 4 and 5~$\mu$m for Designs~A and B, respectively. Table~S3 (Supplement 1) summarizes the implemented dimensions of the focusing input/output grating couplers. The devices were fabricated through the AIM Photonics
Nitride-Only multi-project-wafer (MPW) service on a 300-mm silicon substrate.
Fig.~\ref{fig:device} shows the fabricated chip, representative device
regions, and the measurement setup.

\section{Experimental Results and Discussion}

We characterized the as-fabricated devices without trimming or active
tuning. As shown in Fig.~\ref{fig:device}e, we filtered a visible
supercontinuum source (SuperK FIANIUM, NKT Hamamatsu) with a tunable
filter (SuperK VARIA, NKT Hamamatsu) and coupled the light into the focusing input
grating. We imaged the array emission through a $10\times$ objective
(Mitutoyo M Plan Apo $10\times$/0.28) onto a 2.3-megapixel color CMOS camera (Kiralux CS235CU, Thorlabs).

\begin{figure}[ht!]
\centering
\includegraphics[width=\columnwidth]{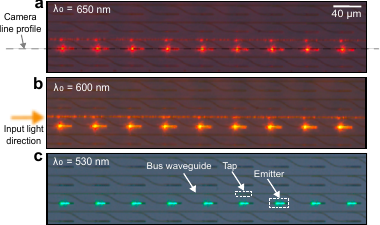}
\caption{Representative camera images of the grating-emitter array under narrowband excitation at \textbf{a}, 650~nm, \textbf{b}, 600~nm, and \textbf{c}, 530~nm. Bright spots indicate individual out-of-plane grating emitters. The dashed line in \textbf{a} marks the line profile used for intensity extraction. Light propagates from left to right. Display levels were adjusted independently for visibility.}
\label{fig:cameraimages}
\end{figure}

The camera field of view captured 15 of the
46 emitters. We selected the
central group, emitters 20--34, to minimize stray light from the input
and residual-output grating couplers. Representative images at 650, 600,
and 530~nm are shown in Fig.~\ref{fig:cameraimages}. For each image, we
extracted a horizontal line profile through the emitter centers
(Fig.~\ref{fig:cameraimages}a, dashed line).

We measure the intensity of each emitter as the background-subtracted peak of its line profile. For each illumination condition we selected the strongest unsaturated RGB channel.
Exposure and gain were kept constant within each image and varied between illumination conditions to ensure linear and unsaturated camera response.
Therefore, we only compare the uniformity of emitters within each condition, and not the absolute intensity between conditions. The measured peak is an end-to-end quantity that includes the tap extraction, the branch transmission, the grating emission, the optical collection, and the camera response.


We quantified uniformity using the coefficient of variation,
$\mathrm{CV}=\sigma/\mu$, calculated over the 15 emitters in a single
frame, where $\sigma$ is the standard deviation and $\mu$ is the mean of
the measured emitter intensities. Because a common multiplicative change
scales both $\sigma$ and $\mu$, common-mode input-coupling fluctuations do
not affect the reported CV. The 95\% bootstrap confidence intervals
quantify spatial sampling uncertainty over the 15 measured emitters rather than
acquisition-to-acquisition measurement repeatability.

\begin{figure}[ht!]
\centering
\includegraphics[width=\columnwidth,keepaspectratio]{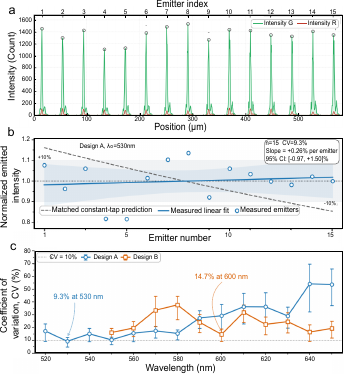}
\caption{Uniformity analysis. \textbf{a}, Intensity profile for Design~A at 530~nm in camera-plane coordinates. Peaks 1--15 correspond to array emitters 20--34; the green camera channel is shown. \textbf{b}, Mean-normalized emitter intensities with a linear fit and matched constant-tap reference ($\alpha_{\mathrm{c}}=0.0217$). Shaded regions indicate $\pm10\%$ about the mean and the 95\% confidence interval of the fitted slope. \textbf{c}, CV versus wavelength; error bars indicate 95\% bootstrap confidence intervals.}
\label{fig:uniformity}
\end{figure}

We measured Designs~A and B from 520 to 650~nm in 10-nm steps.
Figure~\ref{fig:uniformity}a shows the 530-nm intensity profile of
Design~A. Its minimum CV was 9.3\% at 530~nm
(95\% CI, 4.7--12.3\%; $n=15$). Design~B reached a minimum CV of
14.7\% at 600~nm (Fig.~\ref{fig:uniformity}c).

To determine whether the measured uniformity resulted from the designed
tap coupling coefficients rather than simple constant extraction, we compared the
530-nm Design~A profile with the constant-tap reference introduced in
Eq.~\eqref{eq:consttap}. Using
$\alpha_{\mathrm{c}}=\alpha_1=0.0217$, the reference predicts a
mean-normalized slope of $-2.2\%$ per emitter and a CV of 9.5\% over a
15-emitter segment. The predicted CV is close to the measured value of
9.3\%, showing that CV alone cannot distinguish the two tap architectures.

We then characterized the spatial gradient of the cascaded emitter outputs. A linear fit to the
mean-normalized Design~A intensities at 530~nm gave a slope of
$+0.26\%$ per emitter, with a 95\% confidence interval from
$-0.97\%$ to $+1.50\%$. This interval excludes the $-2.2\%$ slope
predicted by the constant-tap reference
(Fig.~\ref{fig:uniformity}b). The measured profile is therefore
inconsistent with monotonic bus depletion and supports the intended
depletion-compensating tap architecture. The remaining emitter-to-emitter
variation can arise from local differences in directional coupling,
branch transmission, grating emission, optical collection, and camera
response.

Each emitter is accompanied by a weak satellite peak. Simulations attribute these satellite peaks to substrate-reflected light that re-emerges through the cladding with a lateral displacement on the order of $10~\mu\mathrm{m}$, comparable to the separation observed experimentally (Supplement 1). These satellite peaks were excluded from the intensity extraction. Laser speckle may also increase the measured emitter-to-emitter variation, such that the reported CV may overestimate
the circuit nonuniformity \cite{shi2025flat}.

The CV varies with wavelength because the directional couplers, grating
emitters, and measurement path are dispersive (Fig.~\ref{fig:uniformity}c). Since each tap has a
different interaction length $L_k$, a wavelength shift perturbs the
coupling coefficients differently along the array. Wavelength-dependent
emission, collection, and camera response further shape the measured
profile, while changes in acquisition settings can affect the noise
level. These contributions are not isolated here.

Under band-integrated illumination from 550 to 650~nm, Design~A produced a CV of 8.8\% over the same 15 emitters. Spectral averaging and reduced speckle contrast likely contribute to the lower value. This end-to-end metric is weighted by the source, optical path, and camera response. Because only emitters 20--34 were imaged, full-array uniformity remains to be experimentally validated.

\section{Conclusion}

We have demonstrated a passive design methodology for distributing visible light uniformly across a 46-channel grating-emitter array. By compensating for progressive bus depletion at the design stage, the approach requires no active phase or amplitude control, trimming, or post-fabrication tuning. The device was fabricated in a commercial 300-mm Si$_3$N$_4$ foundry process and achieved an emitted-intensity CV of 9.3\% at 530~nm across a 15-emitter field of view. The measured spatial gradient excluded the monotonic depletion predicted for a matched constant-tap network and supported the intended variable-tap architecture. Under band-integrated illumination from 550 to 650~nm, the CV was 8.8\%.  We envision that this design methodology can support scalable visible-light delivery for CMOS-matched biosensors, structured illumination, neural interfaces, near-eye displays, and quantum systems requiring parallel optical addressing, including trapped-ion and neutral-atom platforms.



\begin{backmatter}
\bmsection{Funding} This work is supported by the U.S. Army Research Laboratory and the U.S. Army Research Office under contract/grant number W911NF2520055.

\bmsection{Acknowledgment} The authors thank E. Fossum and H. Fang for valuable discussions. 

\bmsection{Disclosures} The authors declare no conflicts of interest.

\bmsection{Data availability} Data underlying the results presented in this paper are not publicly available at this time but may be obtained from the authors upon reasonable request.

\bmsection{Supplementary Information} See Supplement 1 for supporting information.

\end{backmatter}

\bibliography{sample}

\bibliographyfullrefs{sample}

\ifthenelse{\boolean{arxivpreprint}}{\input{supplement}}{}

\end{document}

%% file: supplement.tex
\clearpage
\onecolumn

\setcounter{section}{0}
\setcounter{figure}{0}
\setcounter{table}{0}
\setcounter{equation}{0}
\renewcommand{\thesection}{S\arabic{section}}
\renewcommand{\thefigure}{S\arabic{figure}}
\renewcommand{\thetable}{S\arabic{table}}
\renewcommand{\theequation}{S\arabic{equation}}

\begin{center}
{\large\sffamily\bfseries Supplement 1}\\[6pt]
{\LARGE\bfseries Supporting information for:\\[3pt]
Foundry-Fabricated Silicon Nitride Grating Arrays with\\
Uniform Power Distribution at Visible Wavelengths\par}
\vspace{10pt}

\textsuperscript{1,*}{Amged~Alquliah}, \textsuperscript{1}{Xiaoxin~Wang},
\textsuperscript{1}{Jifeng~Liu}, and \textsuperscript{1,$\dagger$}{Anthony~Rizzo}

\textsuperscript{1}{Thayer School of Engineering, Dartmouth College, Hanover, NH 03755, USA}

\textsuperscript{*}{amged.alquliah@dartmouth.edu} \\
\textsuperscript{$\dagger$}{anthony.j.rizzo@dartmouth.edu}

\end{center}

\vspace{6pt}

\input{supplement_body}

%% file: supplement_body.tex

\section{Bilayer \NoCaseChange{Si$_3$N$_4$} grating-emitter design}
\label{sec:S1}

\begin{figure}[htbp]
\centering
\includegraphics[width=0.8\textwidth]{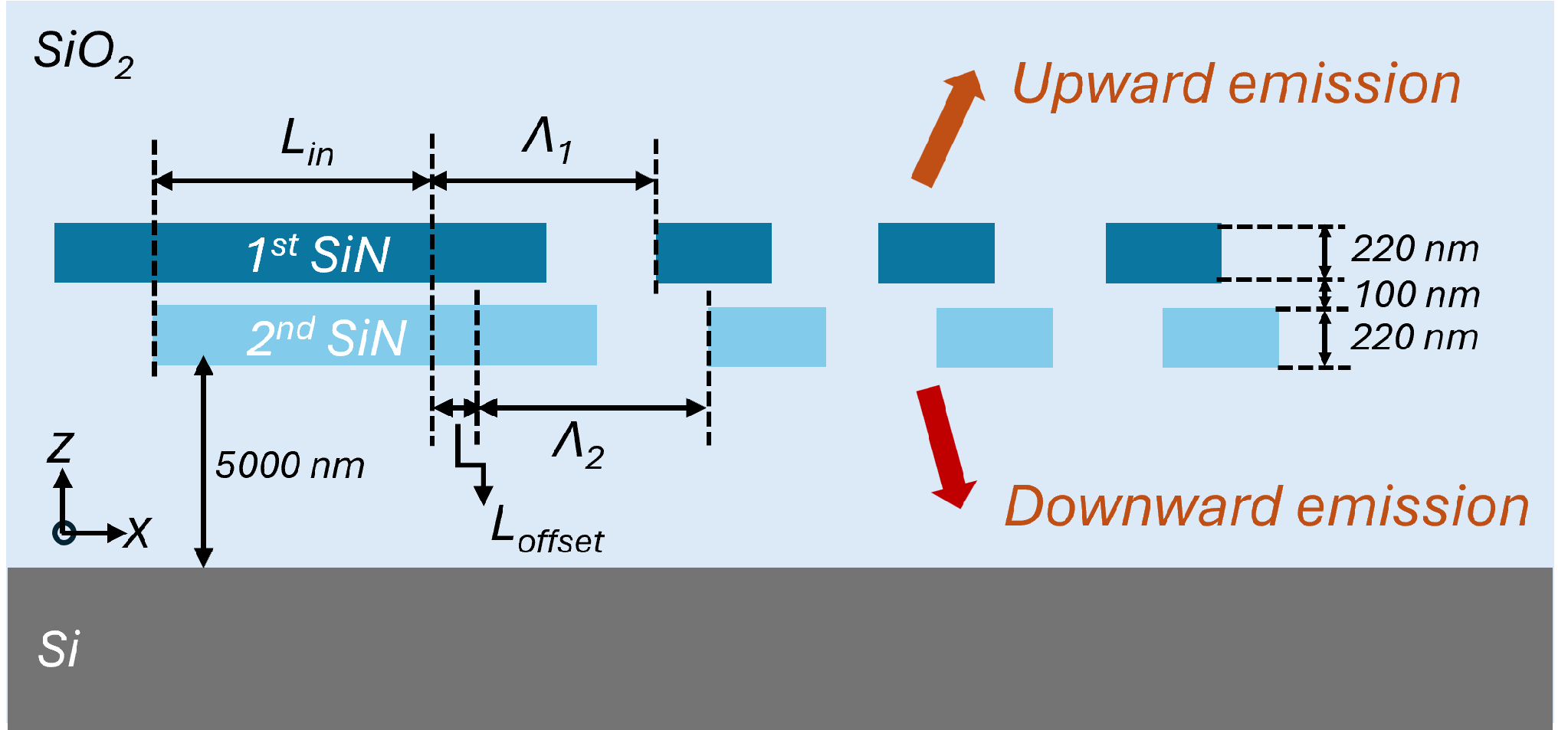}
\caption{Bilayer Si$_3$N$_4$ grating-emitter geometry. Longitudinal cross-section of the dual-layer emitter, adopted from the design approach of Garcia Coleto et al. [S1] and adapted to the AIM Photonics foundry stack. The two 220-nm-thick Si$_3$N$_4$ layers are separated by 100~nm of SiO$_2$ and are patterned with periods $\Lambda_1$ (upper) and $\Lambda_2$ (lower) and fill factor $f$; $L_\mathrm{in}$ is the input taper that feeds the grating from the branch waveguide. The longitudinal offset $L_\mathrm{offset}$ between the upper- and lower-layer perturbations controls the relative phase of the fields scattered by the two layers, so that they add preferentially into the upper half-space and partially cancel downward.}
\label{fig:S1}
\end{figure}

The out-of-plane emitters are bilayer Si$_3$N$_4$ grating structures based on the uniform and unidirectional visible-light grating antennas reported in ref. [S1], adapted here to the target foundry stack (Fig. \ref{fig:S1}). The geometry uses two vertically separated Si$_3$N$_4$ layers with a longitudinal offset between the grating perturbations. This offset sets the relative phase of the fields scattered by the two layers and promotes preferential upward emission without relying on a bottom reflector or an engineered substrate.

The upper and lower Si$_3$N$_4$ layers are each 220~nm thick and separated by 100~nm of SiO$_2$ (Fig.~\ref{fig:S1}). The grating periods of the two layers set the emission angle through the first-order Bragg condition, the fill factor sets the per-period scattering strength and therefore the length over which the guided mode is radiated, the longitudinal offset sets the upward-to-downward scattering ratio, and the input taper couples the branch waveguide into the grating region. The geometries implemented for Designs~A and B are summarized in Table~\ref{tab:S1}. Designs~A and B denote the 550-nm and 650-nm circuit designs of the main text.

\begin{table}[htbp]
\centering
\caption{Implemented grating-emitter geometries for the two circuit designs of the main text. Design~A is the 550-nm circuit design and Design~B the 650-nm circuit design; both use 20 periods and the 220~nm / 100~nm / 220~nm layer stack of Fig.~\ref{fig:S1}.}
\label{tab:S1}
\begin{tabular}{lcc}
\toprule
\textbf{Parameter} & \textbf{Design A (550 nm)} & \textbf{Design B (650 nm)} \\
\midrule
Upper-layer period      & 0.72~$\mu$m & 0.65~$\mu$m \\
Lower-layer period      & 0.75~$\mu$m & 0.65~$\mu$m \\
Fill factor             & 0.51        & 0.50 \\
Longitudinal offset     & 0.36~$\mu$m & 0.88~$\mu$m \\
Number of periods       & 20          & 20 \\
Input taper length      & 4~$\mu$m    & 5~$\mu$m \\
Branch waveguide width  & 380~nm      & 500~nm \\
\bottomrule
\end{tabular}
\end{table}

\section{Numerical simulation of grating emission}
\label{sec:S2}

The bilayer grating emitters were modeled using finite-difference time-domain simulations in Tidy3D [S2], with additional simulations performed in RSoft FullWAVE [S3]. The fundamental TE mode was launched into the emitter taper, and monitors above and below the grating were used to evaluate the upward- and downward-scattered fields. Near-to-far-field projection was then used to calculate the corresponding angular radiation patterns.

\begin{figure}[htbp]
\centering
\includegraphics[width=\textwidth]{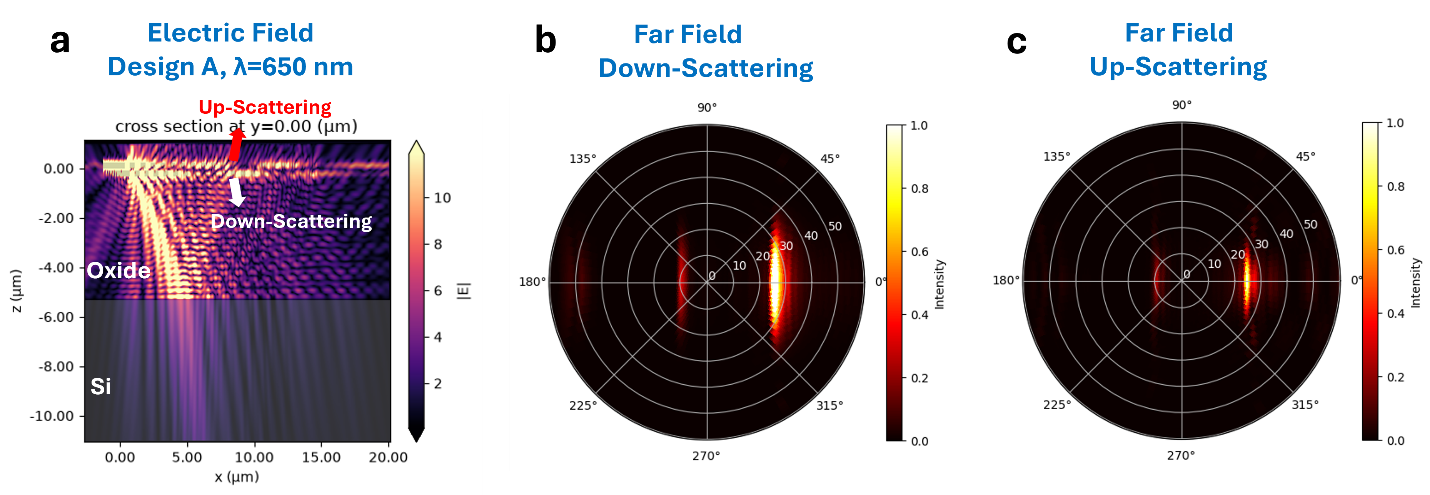}
\caption{Simulated bilayer-grating emission over a thick Si substrate used as the no-return control. The grating geometry shown is evaluated at 650~nm. \textbf{(a)} Simulated electric-field magnitude in the longitudinal cross-section at $y=0$. The bilayer grating produces preferential upward scattering together with a residual downward-scattered lobe propagating at approximately 30$^\circ$ from the surface normal. The downward field enters the thick Si substrate and is mostly absorbed before returning to the grating plane. \textbf{(b,c)} Far-field intensity projected from the downward- and upward-scattered fields, respectively. Both are dominated by localized radiation lobes.}
\label{fig:S2}
\end{figure}

Figure~\ref{fig:S2} shows a representative simulation of the grating geometry at 650~nm. The near field shows preferential upward radiation together with a residual downward-scattered lobe propagating at approximately 30$^\circ$ from the surface normal. In the thick-Si configuration, this downward field is mostly absorbed in the substrate and only a small fraction reflected back to the grating plane. The corresponding upward and downward far-field projections each exhibit a dominant localized radiation lobe.

\section{Wavelength dependence}
\label{sec:S3}

\begin{figure}[htbp]
\centering
\includegraphics[width=\textwidth]{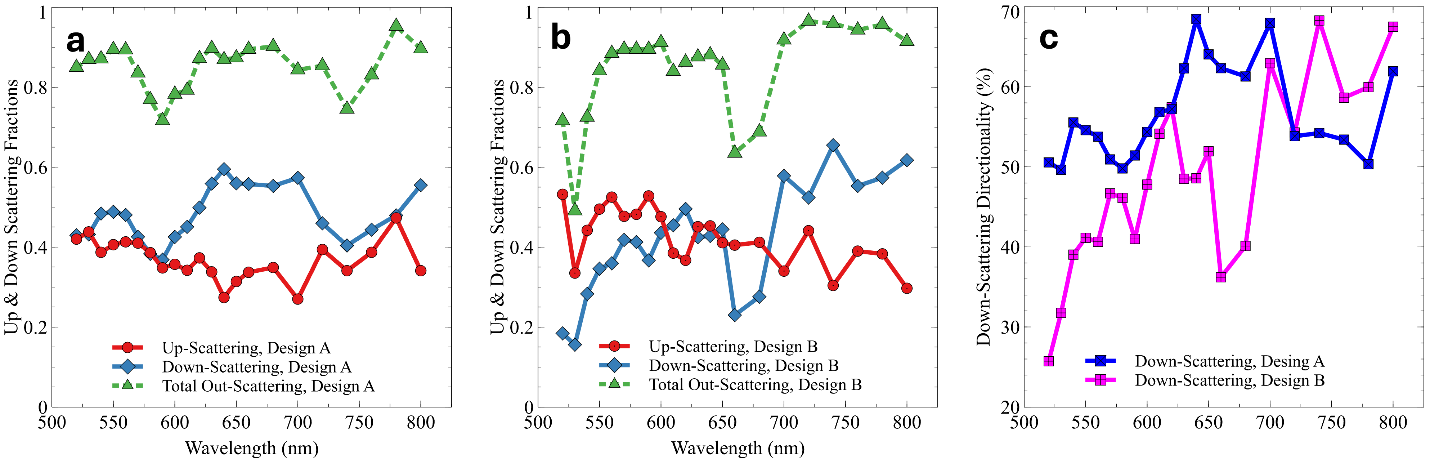}
\caption{Simulated wavelength dependence of the bilayer grating emitters on a 5,000 nm BOX layer on a Si substrate. Upward scattering fraction, downward scattering fraction and total out-scattering of \textbf{(a)} Design~A and \textbf{(b)} Design~B. \textbf{(c)} Comparison of downward scattering directionality of Design~A and Design~B.}
\label{fig:S3}
\end{figure}

The wavelength dependence of the bilayer grating emitters was simulated for a 5,000 nm buried oxide (BOX) layer on a Si substrate. For the thick absorbing-Si substrate, the downward-scattered field is mostly absorbed before it can return to the grating plane. The resulting spectral response therefore primarily reflects the wavelength dependence of the bilayer grating itself, including its phase-matching condition and the relative scattering phase of the two Si$_3$N$_4$ layers. The quantities plotted as upward- and downward-scattered fractions are normalized optical fluxes obtained from the corresponding simulation monitors. Figures~\ref{fig:S3}(a) and \ref{fig:S3}(b) show the upward-scattered fraction, downward-scattered fraction, and total out-scattered fraction as functions of wavelength for Designs~A and B, respectively. The total out-scattered fraction is the sum of the fractions of up-scattering and down-scattering, excluding the light propagating forward or backward in the Si$_3$N$_4$ waveguides. Figure~\ref{fig:S3}(c) compares the down-scattering directionality of Design~A and Design~B, which is defined as the ratio between the downward-scattered power and the total out-scattered power. A large down-scattering directionality indicates a large absorption in the Si substrate or bonded Si sensing chip. A downward-scattering directionality of ${\sim}70\%$ is achieved at $\lambda = 640$~nm for Design~A and at 740~nm for Design~B, respectively. These characteristics of the bilayer Si$_3$N$_4$ emitters can boost the responsivity of a Si sensing chip at the corresponding wavelengths.

\section{Substrate-reflection origin of the secondary peaks}
\label{sec:S4}

\begin{figure}[htbp]
\centering
\includegraphics[width=0.66\textwidth]{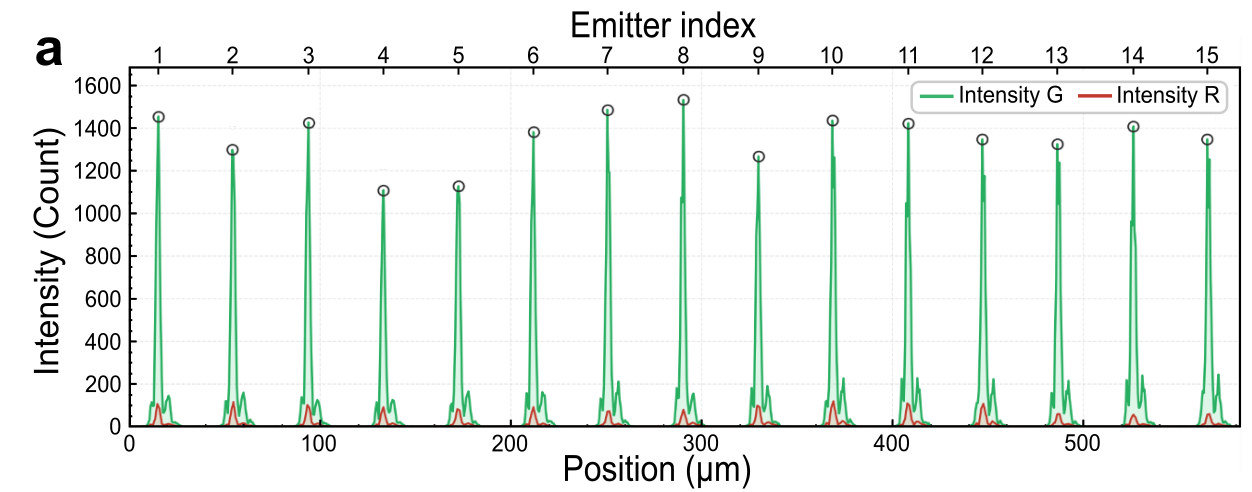}\\[4pt]
\includegraphics[width=0.66\textwidth]{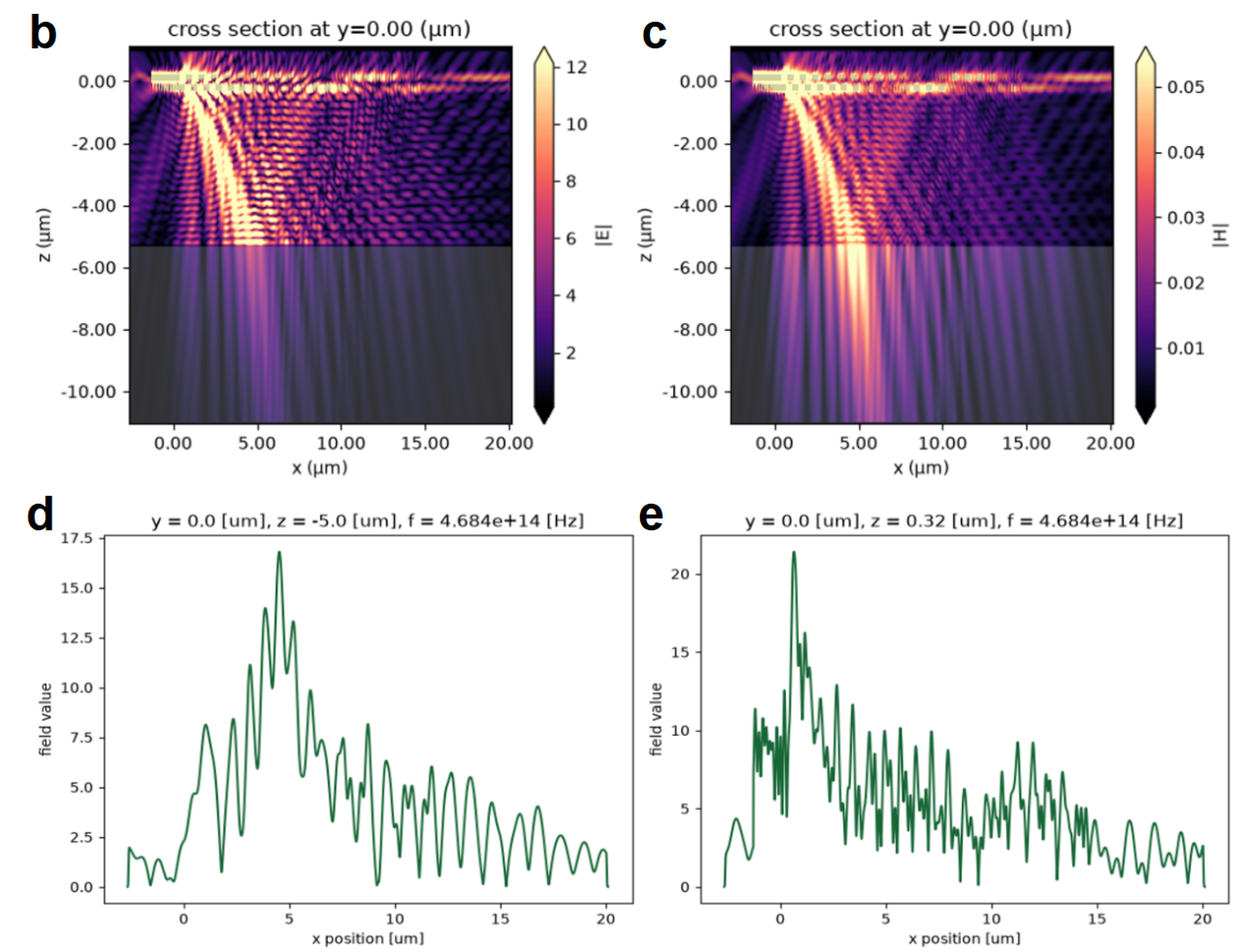}
\caption{Substrate-reflection origin of the secondary emission peaks. \textbf{(a)} Representative measured camera line profile along the emitter row, showing the primary emitter peaks (numbered) and the weaker, laterally displaced secondary features; markers indicate the automatic peak assignment used in the uniformity analysis. \textbf{(b,~c)} Representative Design~A simulation at 640~nm illustrating the reflection mechanism: electric- and magnetic-field magnitudes at $y = 0$ show the downward-scattered light reaching the Si substrate and returning toward the surface with a lateral displacement, producing a standing-wave pattern above the grating. \textbf{(d,~e)} Horizontal cuts of the field magnitude just above the buried Si layer ($z = -3.045~\mu$m) and just above the grating ($z = +0.275~\mu$m) for the same simulation, showing the lateral extent and fringe structure of the returning field.}
\label{fig:S4}
\end{figure}

In the measured camera profiles, each primary emitter peak is accompanied by a weaker, laterally displaced secondary feature (Fig.~\ref{fig:S4}a). To identify its origin, the emitter was simulated using the same foundry stack of the as-fabricated test structure. The mechanism proceeds in four steps. (i) The downward-scattered field first propagates obliquely through the oxide; (ii) it then reflects from the surface of the crystalline-Si substrate, which acts as a partially reflecting layer as well as an absorber; (iii) because of the oblique propagation angle, it returns toward the surface displaced laterally from the emitter by approximately twice the horizontal walk-off accumulated over the round trip; and (iv) the returning field finally interferes with the directly upward-emitted field. The result is a displaced secondary intensity feature consistent with the satellite peaks observed experimentally.

The simulated field maps in Fig.~\ref{fig:S4}(b,~c) show the downward lobe reaching the buried layer and turning back toward the surface, together with the standing-wave pattern that this returning field produces above the grating. The horizontal field cuts of Fig.~\ref{fig:S4}(d,~e), taken just above the buried Si layer at $z = -3.045~\mu$m and just above the grating at $z = +0.275~\mu$m, explicitly show the lateral behavior; both cuts show a strongly modulated envelope that extends several micrometers beyond the region directly beneath the emitter, consistent with a returning beam that is displaced along $x$ and interferes with the directly emitted field. These cuts show the lateral extent and fringe structure of the reflected field more directly than the two-dimensional maps alone.

The simulation supports Si substrate-reflected emission as the origin of the secondary feature rather than the bus-and-tap network. Because the reflected contribution varies with wavelength, the secondary peak amplitude is also wavelength dependent and may contribute to the measured wavelength dependence of the emitter intensity.

\section{Wavelength-resolved near-field maps}
\label{sec:S5}

\begin{figure}[htbp]
\centering
\includegraphics[width=0.62\textwidth]{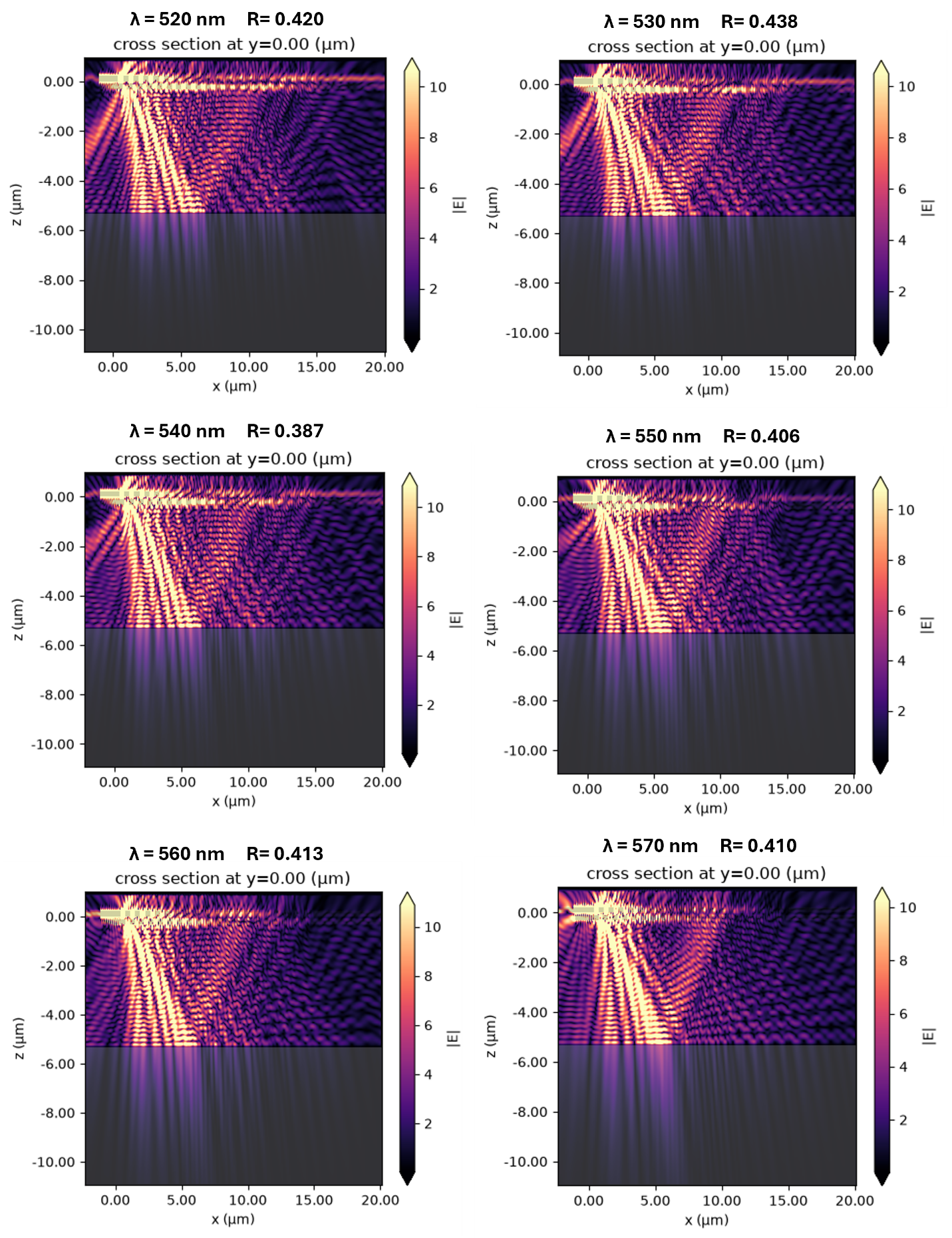}
\caption{Wavelength-resolved near-field maps for Design~A over the foundry stack, 520--570~nm. Each panel shows the electric-field magnitude $|E|$ in the longitudinal cross-section at $y = 0$, with $x$ and $z$ in micrometers, the buried Si layer visible as the horizontal band near $z = -3.4~\mu$m, and the simulated upward-scattered fraction $R$ given above the panel. The fringe pattern above the grating, which is the near-field signature of the returning substrate-reflected beam analyzed in Section~\ref{sec:S4}, changes markedly from one wavelength to the next. These maps show that the mechanism of Fig.~\ref{fig:S4} is present across the measured band rather than at a single wavelength.}
\label{fig:S5}
\end{figure}

\begin{figure}[htbp]
\centering
\includegraphics[width=0.6\textwidth]{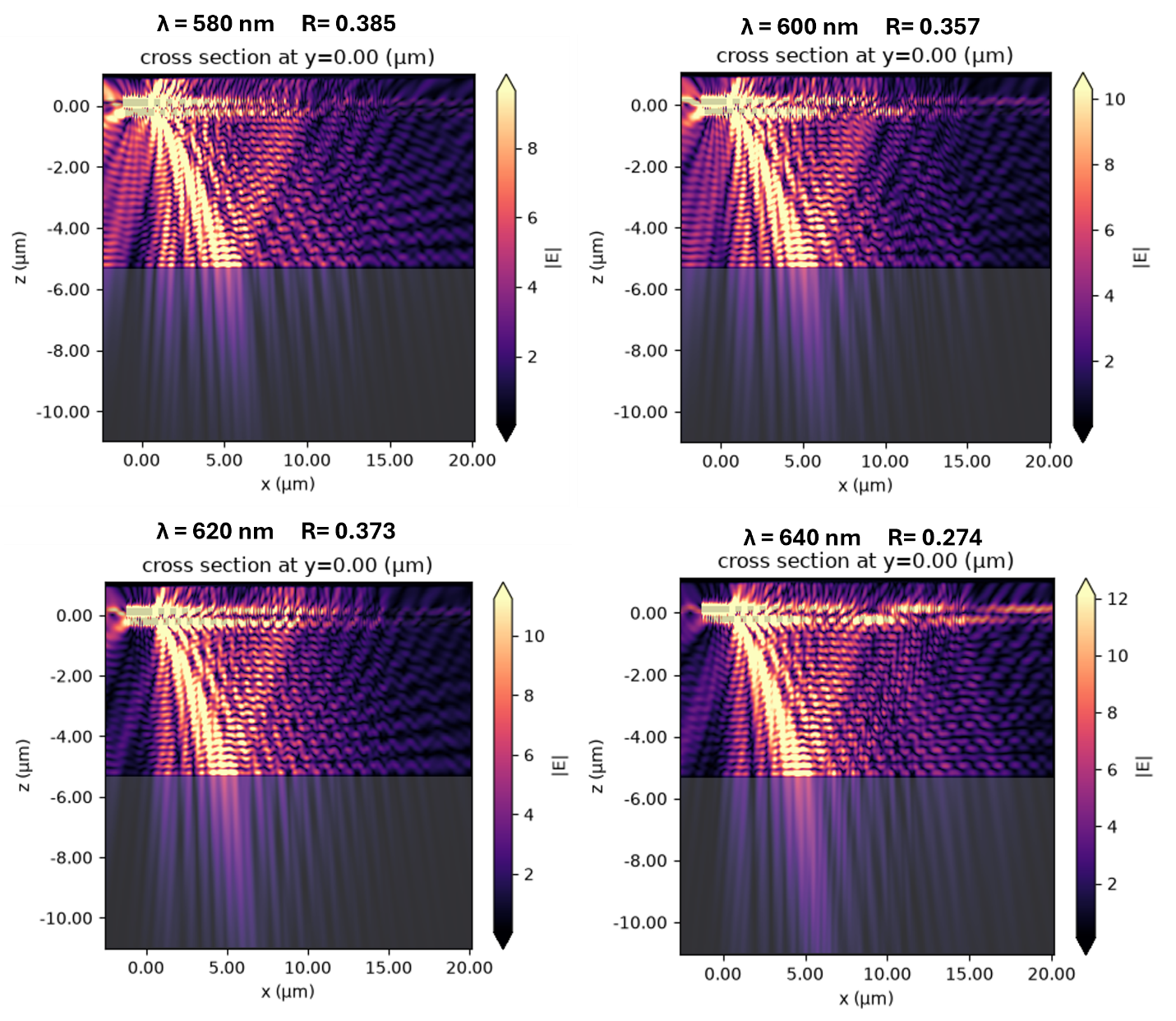}
\caption{Wavelength-resolved near-field maps for Design~A over the foundry stack at 580, 600, 620, and 640~nm, following the same plotting conventions as in Fig.~\ref{fig:S5}.}
\label{fig:S6}
\end{figure}

\begin{figure}[htbp]
\centering
\includegraphics[width=0.62\textwidth]{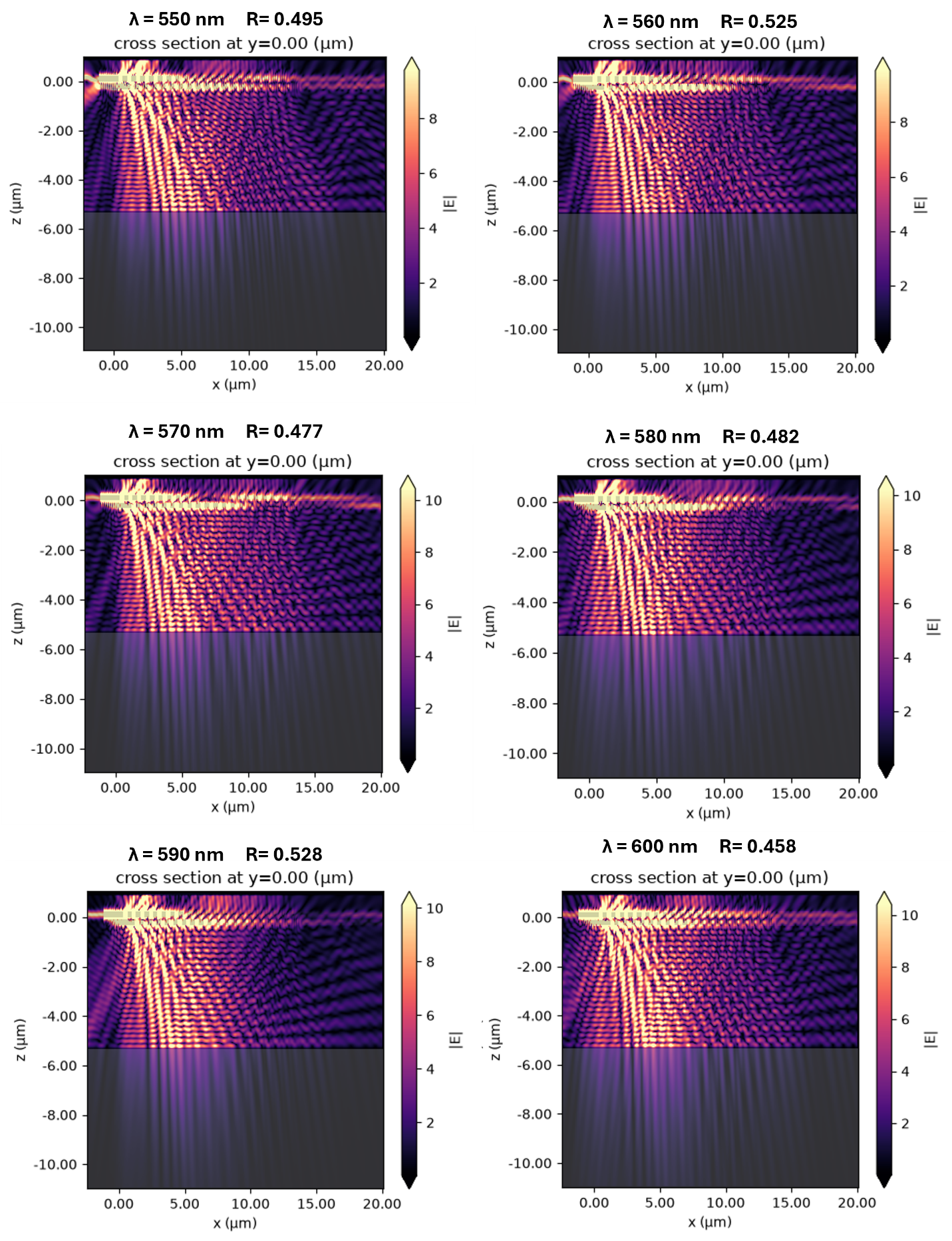}
\caption{Wavelength-resolved near-field maps for Design~B over the foundry stack, 550--600~nm. Design~B has a common 0.65-$\mu$m period in both layers and a 0.88-$\mu$m longitudinal offset (Table~\ref{tab:S1}), so its near-field envelope and its wavelength response differ from those of Design~A at the same wavelength.}
\label{fig:S7}
\end{figure}

\begin{figure}[htbp]
\centering
\includegraphics[width=0.6\textwidth]{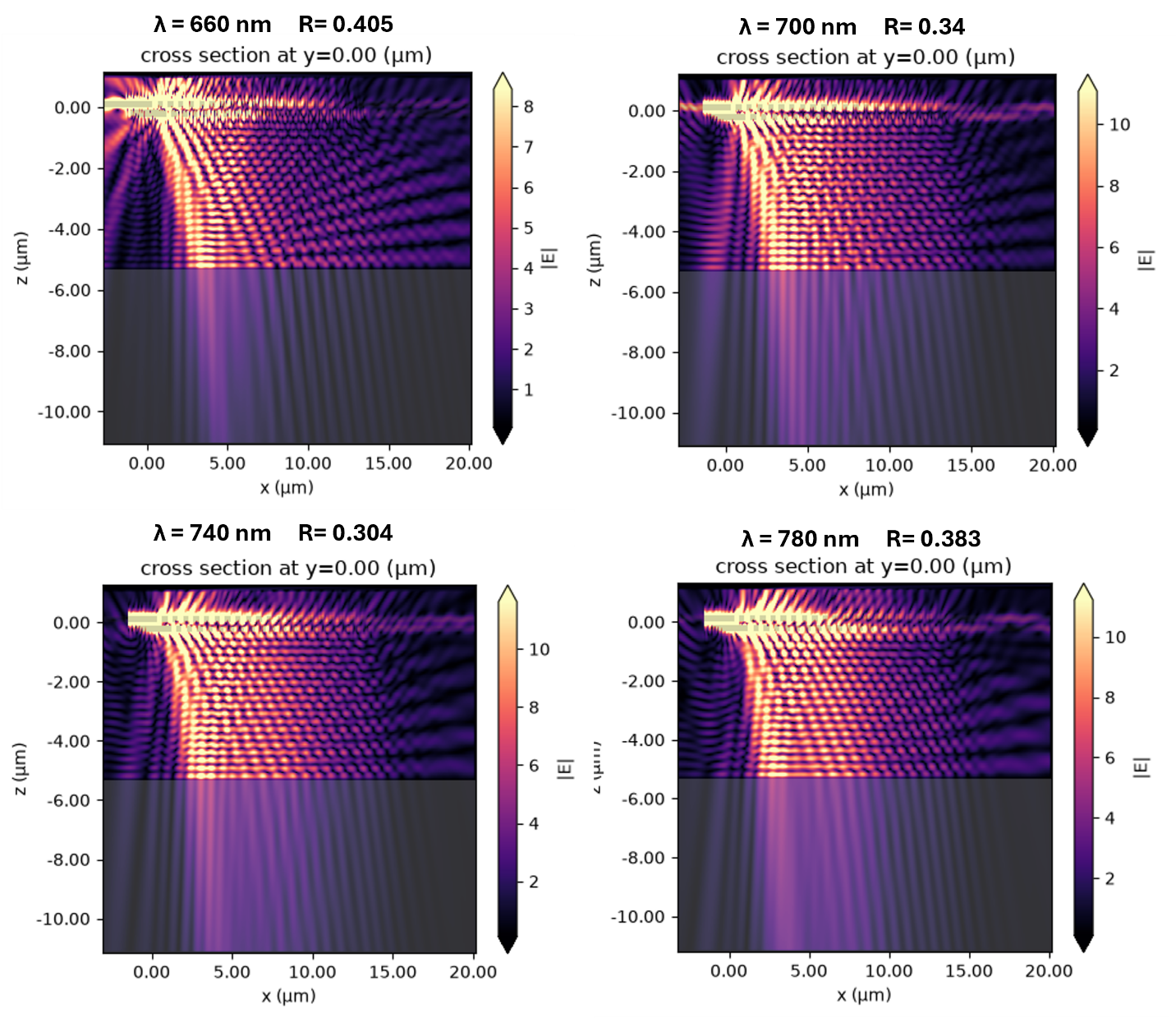}
\caption{Wavelength-resolved near-field maps for Design~B over the foundry stack, 660--780~nm. Across Figs.~\ref{fig:S5}--\ref{fig:S8} the near-field interference structure and the upward-scattered fraction $R$ both vary strongly with wavelength, consistent with wavelength-dependent interference between the directly emitted and substrate-reflected fields.}
\label{fig:S8}
\end{figure}

Figures~\ref{fig:S5}--\ref{fig:S8} show the wavelength-resolved near-field maps that are available for the foundry stack with 5,000 nm BOX on the Si substrate: Design~A at 520, 530, 540, 550, 560, 570, 580, 600, 620, and 640~nm, and Design~B at 550 to 650~nm in 10-nm steps and 660, 700, 740, and 800~nm. These are representative wavelength-resolved simulation results and are part of a complete 520--780~nm sweep for both designs. Each panel is the electric-field magnitude in the longitudinal cross-section at $y = 0$, plotted against the same $x$ and $z$ coordinates and field-magnitude scale conventions, with the buried Si layer visible as the horizontal band near $z = -3.4~\mu$m and the simulated upward-scattered fraction $R$ given above the panel. The near-field interference structure and the upward-scattered fraction $R$ both vary strongly with wavelength, consistent with wavelength-dependent interference between the directly emitted and substrate-reflected fields.

\clearpage 

\section{Directional coupler design details}
\label{sec:S6}
\begin{center}
\captionof{table}{Complete power tap design parameters for the 46-channel bus-and-tap network. For each tap index $k$, the table gives the target power-coupling coefficient $\alpha_k$ and the implemented directional-coupler interaction length $L_k$ for Design~A and Design~B. The listed lengths are those drawn in the layout.}
\label{tab:S2}
\small
\renewcommand{\arraystretch}{0.95}
\setlength{\tabcolsep}{10pt}
\begin{tabular}{ccccc}
\toprule
\textbf{Tap $k$} & \textbf{$\alpha_k$, Design A} & \textbf{$L_k$ ($\mu$m), Design A} & \textbf{$\alpha_k$, Design B} & \textbf{$L_k$ ($\mu$m), Design B} \\
\midrule
1  & 0.02174 & 1.546  & 0.02174 & 1.106 \\
2  & 0.02222 & 1.575  & 0.02222 & 1.126 \\
3  & 0.02273 & 1.605  & 0.02273 & 1.147 \\
4  & 0.02325 & 1.636  & 0.02325 & 1.169 \\
5  & 0.02381 & 1.669  & 0.02381 & 1.191 \\
6  & 0.02439 & 1.703  & 0.02439 & 1.215 \\
7  & 0.02500 & 1.739  & 0.02500 & 1.239 \\
8  & 0.02564 & 1.776  & 0.02564 & 1.265 \\
9  & 0.02631 & 1.815  & 0.02631 & 1.292 \\
10 & 0.02702 & 1.856  & 0.02702 & 1.320 \\
11 & 0.02778 & 1.899  & 0.02777 & 1.350 \\
12 & 0.02857 & 1.944  & 0.02857 & 1.381 \\
13 & 0.02941 & 1.991  & 0.02941 & 1.414 \\
14 & 0.03030 & 2.042  & 0.03030 & 1.448 \\
15 & 0.03125 & 2.094  & 0.03125 & 1.485 \\
16 & 0.03225 & 2.150  & 0.03225 & 1.523 \\
17 & 0.03333 & 2.209  & 0.03333 & 1.564 \\
18 & 0.03448 & 2.272  & 0.03448 & 1.607 \\
19 & 0.03571 & 2.338  & 0.03571 & 1.653 \\
20 & 0.03703 & 2.409  & 0.03703 & 1.702 \\
21 & 0.03846 & 2.484  & 0.03846 & 1.754 \\
22 & 0.04000 & 2.564  & 0.03999 & 1.810 \\
23 & 0.04166 & 2.650  & 0.04166 & 1.869 \\
24 & 0.04347 & 2.743  & 0.04347 & 1.933 \\
25 & 0.04545 & 2.842  & 0.04545 & 2.001 \\
26 & 0.04761 & 2.948  & 0.04761 & 2.075 \\
27 & 0.04999 & 3.064  & 0.04999 & 2.155 \\
28 & 0.05262 & 3.189  & 0.05262 & 2.242 \\
29 & 0.05555 & 3.324  & 0.05554 & 2.336 \\
30 & 0.05881 & 3.472  & 0.05881 & 2.439 \\
31 & 0.06249 & 3.634  & 0.06248 & 2.551 \\
32 & 0.06665 & 3.811  & 0.06665 & 2.675 \\
33 & 0.07141 & 4.007  & 0.07141 & 2.811 \\
34 & 0.07690 & 4.222  & 0.07690 & 2.961 \\
35 & 0.08331 & 4.462  & 0.08331 & 3.129 \\
36 & 0.09088 & 4.728  & 0.09088 & 3.316 \\
37 & 0.09997 & 5.025  & 0.09996 & 3.525 \\
38 & 0.11107 & 5.357  & 0.11106 & 3.760 \\
39 & 0.12495 & 5.727  & 0.12494 & 4.024 \\
40 & 0.14279 & 6.138  & 0.14278 & 4.321 \\
41 & 0.16658 & 6.587  & 0.16656 & 4.652 \\
42 & 0.19988 & 7.067  & 0.19984 & 5.022 \\
43 & 0.24981 & 7.574  & 0.24975 & 5.449 \\
44 & 0.33299 & 8.245  & 0.33289 & 6.089 \\
45 & 0.49923 & 10.728 & 0.49900 & 8.345 \\
46 & 0.99692 & 22.406 & 0.99600 & 18.874 \\
\bottomrule
\end{tabular}
\end{center}

\clearpage

\section{Fiber-chip grating coupler design details}
\label{sec:S7}

\begin{center}
\captionof{table}{Implemented focusing fiber-chip grating-coupler designs. Both couplers are fully etched in the 220-nm upper Si$_3$N$_4$ layer. The design wavelength is the wavelength at which the coupler was optimized in Lumerical; the tooth width and gap follow from the period and duty cycle; the number of periods follows from the target grating length; and $n_\mathrm{eff}$ is the effective index supplied to the layout generator to set the focusing curvature.}
\label{tab:S3}
\begin{tabular}{lcc}
\toprule
\textbf{Parameter} & \textbf{Coupler design 1} & \textbf{Coupler design 2} \\
\midrule
Grating period                  & 482.2~nm            & 571~nm \\
Duty cycle                      & 0.484               & 0.480 \\
Tooth width                     & 233.4~nm            & 274.1~nm \\
Gap                             & 248.8~nm            & 296.9~nm \\
Target grating length           & 9~$\mu$m            & 12~$\mu$m \\
Number of periods               & 19                  & 22 \\
Si$_3$N$_4$ thickness           & 220~nm              & 220~nm \\
Etch depth                      & 220~nm (full etch)  & 220~nm (full etch) \\
Focusing taper length           & 10~$\mu$m           & 10~$\mu$m \\
Taper opening angle             & 40$^\circ$          & 40$^\circ$ \\
Fiber incidence angle           & 10$^\circ$          & 7$^\circ$ \\
Cladding index, $n_\mathrm{clad}$ & 1.457             & 1.454 \\
\bottomrule
\end{tabular}
\end{center}
\vspace{12pt}

\section*{Funding}
This work is supported by the U. S. Army Research Laboratory and the U. S. Army Research Office under contract/grant number W911NF2520055.

\makeatletter
\@ifundefined{bibnumfmt}{\renewcommand{\@biblabel}[1]{[S#1]}}{\renewcommand{\bibnumfmt}[1]{[S#1]}}
\makeatother